\documentclass[conference]{IEEEtran}
\makeatletter
\def\ps@headings{%
\def\@oddhead{\mbox{}\scriptsize\rightmark \hfil \thepage}%
\def\@evenhead{\scriptsize\thepage \hfil \leftmark\mbox{}}%
\def\@oddfoot{}%
\def\@evenfoot{}}
\makeatother
\usepackage{graphicx,epstopdf, amsmath,epsfig,subfigure,endnotes,footnote,multicol,sidecap,url,colortbl,booktabs,algorithm,algorithmic,multirow,enumerate}
\usepackage{amsfonts}
\usepackage [english]{babel}
\usepackage [autostyle, english = american]{csquotes}
\MakeOuterQuote{"}

\usepackage{balance}

\usepackage{amsthm}
\usepackage{mathtools}

\usepackage{environ}
\theoremstyle{plain}

\usepackage{color}
\definecolor{red}{rgb}{1,0,0}

\addto\captionsenglish{\renewcommand{\figurename}{Fig.}}

\NewEnviron{myequation}{%
    \begin{equation}
    \scalebox{0.8}{$\BODY$}
    \end{equation}
    }

\begin{document}
\title{Cyber Security in Smart Manufacturing (Threats,
Landscapes \& Challenges)}

\author{
Rahat Masum\\
Department of Computer Science,\\
Tennessee Tech University, Cookeville, TN, USA\\
rahatapu103@gmail.com

}

\maketitle
 
\begin{abstract}
Industry 4.0 is a blend of the hyper-connected digital industry within two world of Information Technology (IT) and Operational Technology (OT). With this amalgamate opportunity, smart manufacturing involves production assets with the manufacturing equipment having its own intelligence, while the system-wide intelligence is provided by the cyber layer. However Smart manufacturing now becomes one of the prime targets of cyber threats due to vulnerabilities in the existing process of operation. Since smart manufacturing covers a vast area of production industries from cyber physical system to additive manufacturing, to autonomous vehicles, to cloud based IIoT (Industrial IoT), to robotic production, cyber threat stands out with this regard questioning about how to connect manufacturing resources by network, how to integrate a whole process chain for a factory production etc. Cybersecurity confidentiality, integrity and availability expose their essential existence for the proper operational thread model known as digital thread ensuring secure manufacturing. In this work, a literature survey is presented from the existing threat models, attack vectors and future challenges over the digital thread of smart manufacturing.

\end{abstract}



\begin{IEEEkeywords}
smart manufacturing; cybersecurity; survey; digital thread; artificial intelligence.
\end{IEEEkeywords}

\section{Introduction}
\label{sec:Intro}

Smart manufacturing is the modern automation of industrial process. It is the box of combination where cyber space comes into contact with the machine domain. The human-based production process goes through the change by the uprising manner of smart industry. Every factory is now connected with some public network or internal communication system to make the automated process of production. This system is controlled by software enabled modules or control systems which will pass the required parameters to the physical machine via some logical organization of program set. Software can be compromised by malicious code, such as Trojans, viruses, and runtime attacks. Communication protocols are subject to protocol attacks, including man-in-the-middle and denial-of-service attacks. According to NIST (National Institute of Standards and Technology), physical and logical layers respond in real time to meet the changing demands and conditions in the factory, supply network, and customer needs. NIST Framework is a tool which analyzes the risks of a particular organization neutrally. It analyzes and applies the standards which are applicable in risk evaluation for that organization. In general, it lays out a method of risk analysis framed by standards and best practices, so any organization can use it.

To provide a comprehensive description of existing smart manufacturing, in this work, we try to identify key threats and challenges regarding cybersecurity scopes in the manufacturing. The rest of the paper is organized as in Section II, we discuss some aspects of Industry 4.0. Basic structures of smart manufacturing are overviewed in Section III. In Section IV, regular practices for Cybersecurity is defined with the help of CIA Triad. The literature survey on the  scope of threats and existing scenarios for manufacturing domain is presented on Section V according to the proposed taxonomy. Section VI discusses some summary over the review and exiting challenges for better solutions. Finally Section VI holds the conclusion and future direction of our works.





\section{Industry 4.0}
\label{sec:I4}

Industry 4.0 as a convergence of two worlds which have been disconnected thus far: Information Technology (IT) and Operational Technology (OT). The bridging of digital and physical, cyber-physical production systems and the Industrial IoT. Industrial Data Space, which aims to define the data architecture for the connection of smart services and IoT in a landscape of ever more data wants to become a global standard. Industrial Data Space focuses on the connection of several existing platforms and wants to enable partners to share data for digital transformation purposes within a clear pre-defined model whereby data sovereignty and data security are among the key focus areas.\\

Fourth wave of technological advancement is known as Industry 4.0~\cite{russmann2015industry}. It introduces this new digital technology as a transformation which is based on nine foundational advances. In this transformation, sensors, machines, workpieces, and IT systems will be connected along the value chain beyond a single enterprise. These connected systems (also referred to as cyber physical systems) can interact with one another using standard Internet-based protocols and analyze data to predict failure, configure themselves, and adapt to changes. The authors describe Industry 4.0 case studies from German industrial automation as to gather and analyze data across machines, enabling faster, more flexible, and more efficient processes. The nine pillars of the advancement that transforming Industrial Production are listed as:
\begin{itemize}
\item  Big Data and analytics
\item  Autonomous Robots
\item  Simulation
\item  Horizontal and Vertical System Integration
\item  The Industrial Internet of Things(IIoT)
\item  Cybersecurity
\item  The Cloud
\item  Additive Manufacturing
\item  Augmented Reality
\end{itemize}

Russmann \textit{et al.} analyzed the outlook for manufacturing in Germany for the Industry 4.0 impact. They have mentioned the effect of manufacturing over producers’ entire value chain, from design to after-sales service and found the benefits in four areas as:
\begin{itemize}
\item Productivity
\item Revenue Growth
\item Employment
\item Investment
\end{itemize}

As manufacturers demand the greater connectivity and interaction of Industry 4.0 capable machines and systems in their factories, manufacturing-system suppliers will have to expand the role of IT in their products.\\

Manufacturing companies are facing impact with Cyber-Physical Systems which introduced as a fusion of the physical and the virtual worlds, the Internet of Things and the Internet of Services. Based on connectivity and computing power, manufacturing equipment will turn into CPPS, Cyber-Physical Production Systems - software enhanced machinery, also with their own computing power, leveraging a wide range of embedded sensors and actuators, beyond connectivity and computing power~\cite{almada2016industry}.


\section{Smart Manufacturing}
\label{sec:SM}
Smart Manufacturing depends on the scope and degree of automation of a manufacturing floor and the integration of various functional production areas. According to Kusiak \textit{et al.}, smart manufacturing integrates manufacturing assets of today and tomorrow with sensors, computing platforms, communication technology, data intensive modeling, control, simulation and predictive engineering. Smart manufacturing utilities the concepts of the cyber-physical systems, Internet of things (and everything), cloud computing, service-oriented computing, artificial intelligence and data science. The author uses different terms to describe automated manufacturing since 1980s, ranging from flexible manufacturing cells and flexible manufacturing systems to computer integrated manufacturing and intelligent manufacturing~\cite{kusiak2017smart}. There is a presentation of two basic layers linked by the interface as:
\begin{itemize}
\item Manufacturing equipment layer
\item Cyber layer
\end{itemize} 
The manufacturing equipment has its own intelligence, while the system-wide intelligence is provided by the cyber layer. The whole smart manufacturing concept is based on six components as mentioned by the author:
\begin{itemize}
\item Manufacturing technology and processes
\item Materials
\item Data
\item Predictive engineering
\item Sustainability
\item Resource sharing and networking
\end{itemize}
Lu \textit{et al.} talk about one challenge for the growing volumes and reliance on data to make cyber security paramount to the progress in smart manufacturing and business competitiveness. This is especially important as data assets will become a growing indicator of the market value of a company. The increasing degree of automation and system autonomy will raise the importance of human and machine safety~\cite{lu2016current}. Condition monitoring solutions and warning systems will gain importance. In fact, the commonality between solutions developed for equipment diagnosis and cyber security may be explored.\\

Direct Digital Manufacturing (DDM) plays significant role in the mainstream supply chain of manufacturing industry which produces parts from a Computer Aided Design drawing (CAD) directly. Due to potential economic and security implications of DDM, one challenge of Industry 4.0 is to address cyber security risks in timely way and develop standards, systems and processes for security before wide scale adoption of the technology limits, or the deployment of protection mechanisms~\cite{DDM}. As per the definition discussed by Glavach \textit{et al.} DDM uses 3D computer-aided design files to drive computer controlled fabrication of parts~\cite{glavach2017applying} .\\

In DDM new hardware and software driven systems use variety of components and ingredients to build up layers of materials to create complex structures which are designed, mode
led and tested within the cyber world prior to manufacturing.\\

NIST defines Smart Manufacturing as 'fully-integrated and collaborative manufacturing systems that respond in real time to meet the changing demands and conditions in the factory, supply network, and customer needs'~\cite{nist}.\\

Some literature servey was done over smart manufacturing historical background where Kang \textit{et al.} talk about NIST provided details. NIST established the strategies of dynamic production system and rapid design-to-product through smart manufacture research programs~\cite{kang2016smart}. It suggested three key action technologies, namely, decentralized control network, digital manufacturing, decentralized machine intelligence.

\subsection{Digital Thread}
The digital thread refers to the communication framework of a connected data flow with an integrated view of the asset’s data \textit{Life} cycle and \textit{Functional} perspectives. Along the path from design to manufacture, a lot of information gets passed from computer to computer and machine to machine.\\

A discussion has been presented as ‘digital thread’ defined by the smart manufacturing on the basis of Industry 4.0. In a factory floor system since the era of skilled machinist operating from handmade paper drawn product designs has now converted into the digital world with the fourth revolution, digital thread refers to the networks of computers, automated machines, ubiquitous sensors, and technicians whose job is to convert digital data into physical parts and assemblies. Design, manufacturing and product support operations are driven by a digital thread of technical data and can be shared throughout the supply chain~\cite{national2014cybersecurity}.\\

For different industries in the smart manufacturing sector digital thread contains layer of levels based on the product generated. In this paper we will present a generalized high level representation of that.\\
\begin{figure}[!h]
\includegraphics[width=0.5\textwidth]{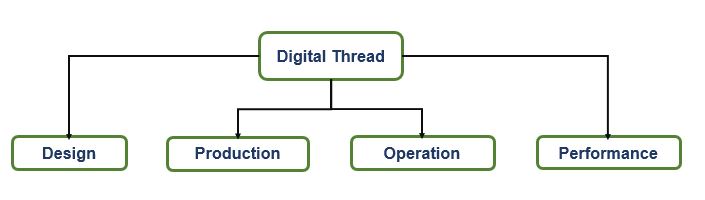}
\caption{\label{fig:}Digital thread}
\end{figure}

The digital thread should provide a formal framework for controlled interplay of authoritative technical and as-built data with the ability to access, integrate, transform and analyze data from disparate systems throughout the product life-cycle into actionable information~\cite{digital}. The product life-cycle includes: Design, Procurement, Test \& Evaluation, Production, Field Operation, and Sustainment Services. The digital thread and digital twin include as-designed requirements, validation and inspection records, as built data, as-flown data, and as-maintained data.\\

From design, to simulation, to build plan, to process monitoring and control, to verification standards organizations, such as ASTM (American Society for Testing and Materials), have already begun to establish file formats that address some of these data links but additional data formats must be established to realize the greatest potential of cyber-enabled manufacturing~\cite{nassar2013proposed}. Ideally, data necessary for part design, manufacturing, qualification and testing should be part of a single digital thread and easily accessible, traceable and inter-operable with all machines along the process chain.\\

\section{Cyber Security Practices}
\label{sec:CSP}
For the recent years, cyber attacks are growing with a mentionable occurrences all over the world from industrial manufacturing to factory production systems. Smart manufacturing being a part of Industry 4.0, now becomes one of the prime targets of cyber threats due to vulnerabilities in the existing process of execution. 
\subsection{\textbf{Threats}}
As of the discussion regarding Digital thread in the manufacturing industries the whole cyber space can be grouped into some domains if someone wants to specify the threat model for that. The digitalized transformation leads to the following domains in the chain:
\begin{itemize}
\item \textbf{Network:} No smart industrial system can be thought of the network connectivity both in factory inside and outside public accesses. The common practices for securing this factory floor network is the firewall protection mechanism to allow or restrict access according to the privilege rights defined by the industry management. However, lack of proper configuration or maintenance planning factory network is always in the threat of cyber intrusion for any industry in the process chain.\\

\item \textbf{Database:} Digital Thread holds the information of different product designs, process parameters, production tracking, customer feedback, sales history etc. in the form of some representational manner of digital data. For storing this huge chunk of sensitive information each smart industry usually has more than one database systems. With logical and physical  encryption protected practices, these databases are always a lucrative link for attackers to gain all the information in hand. As a result, threat rates for database attack is very high in the common protection plan.\\

\item \textbf{Smart Devices:} Due to rapid growth of IoT (Internet of things) devices and cloud based platform services, single sign-on is one of the most convenient method of access management in the factory floor. Everyday workers as well as other employees such as managers, sales person, suppliers get connected to the internal and external network of the manufacturing domain with respective purposes through the smart devices. These device can be used as a strong link to get access to factory production machines and other system units. Cyber space of the connected device network channel cannot be monitored continuously due to high maintenance cost and time consuming manner of the mechanism. Therefore, vulnerabilities will increase with these IoT connectivities if not addressed properly.\\

\item \textbf{Cloud Manufacturing:} Manufacturing enterprises handle a wide range of sensitive data between customers, suppliers and equipment vendors. Keeping this feature in mind, industries are extending their network and started processing by utilizing well established manufacturing resources with Enterprise Resource Planning (ERP) through the cloud where cloud is a network that allows to store and manage data, files, and software over the Internet. Maintaining 24-hour security to watch out for any potential security breaches and fraudulent code in the cloud now becomes new responsibilities to the administrative. Cyber threat stands out with this regard questioning about how to connect those distributed manufacturing resources by network, how to integrate resource so that a manufacturing task can be finished accordingly and also transaction or circulation of on-demand use of manufacturing resource and capability.
\end{itemize}
\subsection{\textbf{CIA Triad}}
Cyber threats to manufacturing enterprises may be motivated by espionage, financial gain or other reasons to compromise data.

\begin{figure}[!h]
\includegraphics[width=0.5\textwidth]{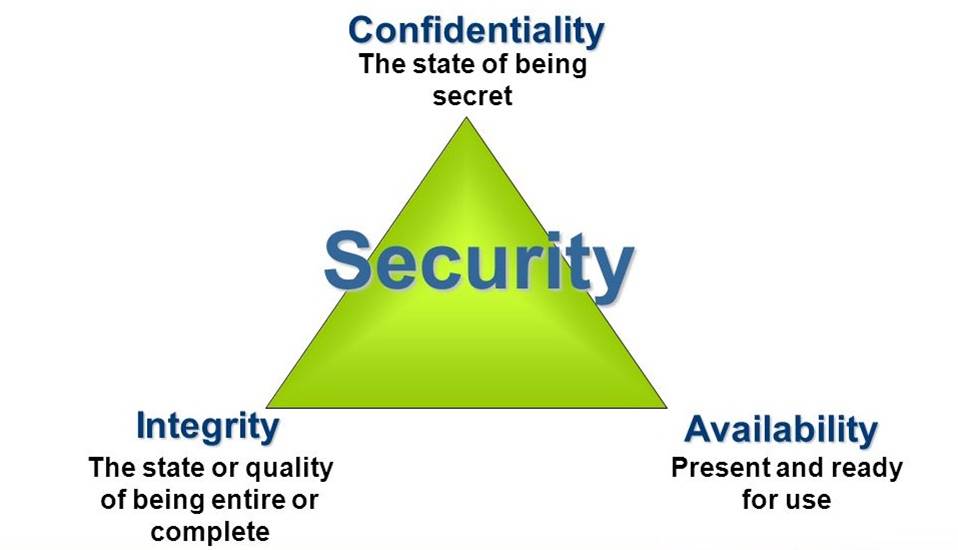}
\caption{\label{fig:}CIA Triad~\cite{cia}}
\end{figure}

For the advanced manufacturing enterprise, these concerns are translated~\cite{national2014cybersecurity} as:

\begin{itemize}
\item Theft of technical data, including critical national security information and valuable commercial intellectual property as a \textit{Confidentiality} concern.

\item Alteration of data, thereby altering processes and products as an \textit{Integrity} concern.

\item Impairment or denial of process control, thereby damaging or shutting down operations as an \textit{Availability} concern.
\end{itemize}

These concerns exist from the point of creation of the technical data, through its access at any point in the supply chain, to its use to control physical manufacturing processes throughout the product life cycle.

Cebula \textit{et al.} discuss the sources of operational cybersecurity risk into different classes~\cite{cebula2010taxonomy}:
\begin{itemize}
\item Actions taken by people either deliberately or accidentally that impact cyber security as: fraud, sabotage (carried out against targeted key assets by someone possessing or with access to inside knowledge), theft (the intentional, unauthorized taking of organizational assets). This action violates the \textit{Confidentiality} of the system.
\item Failed internal processes represents the ability to implement, manage, and sustain cyber security, such as process design, execution, and control, inadequate or nonexistent ability to escalate abnormal or unexpected conditions for action by appropriate personnel violating \textit{Integrity} characteristics
\item Systems and technology failures represent inability to handle a given load or volume of information, inability to complete instructions or process information within acceptable parameters (speed, power consumption, heat load, etc.), changes made to the application or its configuration by a process lacking appropriate authorization making the \textit{Availability} property violated.
\end{itemize}

\section{Cybersecurity Scope in Smart Manufacturing}
\label{sec:TX}
The risk of cyber attacks directed at ICS-based manufacturing infrastructures and processes is a great concern to companies who produce goods, particularly those made for public sconsumption~\cite{ics}. NIST recognizes this concern and is working with industry to solve these challenges through the implementation of cybersecurity technologies. In addition to this challenge, NIST provides the CSF (Cyber Security Framework) for any manufacturing entity interested in enhancing the security of its infrastructure.
\subsection{\textbf{Taxonomy}}
In this research, a brief survey approach has been done to find out the existing scenarios for cyber domains in the smart manufacturing industry. Since smart manufacturing covers a vast area of production industries from cyber physical system to additive manufacturing, to autonomous vehicles, to cloud based IIoT ( Industrial IoT), to robotic production, a taxonomy has been developed for convenience division.\\
\begin{figure}[!h]
\includegraphics[width=0.5\textwidth]{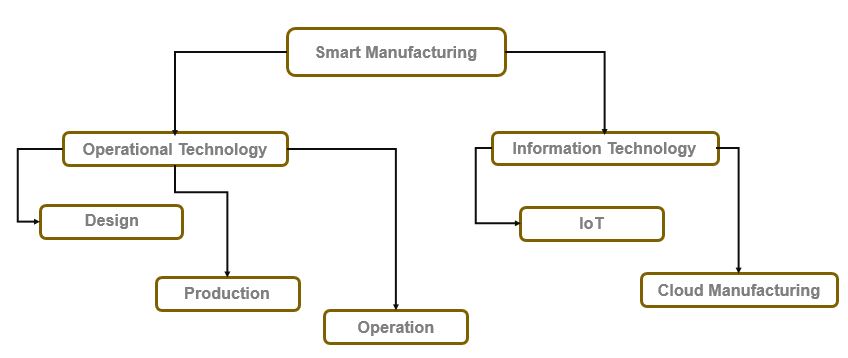}
\caption{\label{fig:}Proposed taxonomy}
\end{figure}

As discussed previously, Digital Thread goes through the OT and IT whole chain, our analysis will be done over the OT sector with the scope of cybersecurity breaches and possible protection and solution mechanisms where applicable.

\subsection{Operational Technology Landscape}

Traditional data security practices are not much helpful if any hardware or software component of the system is built to send the inner data to outside of the system or leave the back door open for intruders intentionally. Devices can be trusted only if the chips are free of hidden malicious circuits which may be inserted during the design or manufacturing process of the chips~\cite{pal2017cyber}.\\ 

In this literature survey different threats and possible/proposed solutions are grouped according to the three main key aspects of cybersecurity. For the convenience of discussion, existing works have grouped into three subsectors of OT process. Based on the threat model, existing works are discussed as follows. \begin{table}[h]

\begin{tabular}{|p{0.3\linewidth}|p{0.2\linewidth}|p{0.3\linewidth}|}
\hline
\textbf{Author} & \textbf{OT Sector} & \textbf{CIA Triad} \\ \hline
Wong \textit{et al.} & Design & Confidentiality \\ \hline
Yampolskiy \textit{et al.} & Design & Confidentiality \\ \hline
Zeltmann \textit{et al.} & Design & Integrity\\ \hline
Sturm \textit{et al.} & Design & Integrity\\ \hline
Straub \textit{et al.} & Design &  Integrity \\ \hline
Vincent \textit{et al.} & Design & Integrity \\ \hline
Chhetri \textit{et al.} & Design & Integrity \\ \hline
Sadeghi \textit{et al.} & Design & Availability\\ \hline
Xun Xu & Production & Confidentiality \\ \hline
Pal \textit{et al.} & Production & Confidentiality \\ \hline
Moore \textit{et al.} & Production & Integrity \\ \hline
Khan \textit{et al.} & Production & Availability \\ \hline
Bayens \textit{et al.} & Production & Integrity \\ \hline
Sadeghi \textit{et al.} & Operation & Confidentiality \\ \hline
Zhang \textit{et al.} & Operation & Confidentiality \\ \hline
Moore \textit{et al.} & Operation & Availability \\ \hline
Pal \textit{et al.} & Operation & Availability \\ \hline

\end{tabular}
\end{table}

\subsection{Confidentiality}
\subsubsection{\textbf{Design}}
Security and privacy are two distinct but related issues in the area of information technology. The notion of security concerns preventing unauthorized access to proprietary data, while privacy deals with considerations such as who own the data, who can access it, and how it can be used. Attackers often target essential and highly sensitive information, such as product designs, financial data, marketing plans, customer and supplier lists, and partnership agreements. The loss of sensitive information and other forms of enterprise information can lead to significant financial or business losses.\\

In the manufacturing context, data is considered sensitive due to various aspects of industrial operation, including highly sensitive information about products, business strategies, and companies. Sharing data with internal departments or external vendors requires secure mechanisms (e.g., access control infrastructure) to prevent information leakage. Sensitive data, which includes blueprints, manufacturing processes, cost information and operational data, should be protected.\\

Wong \textit{et al.} talk about several key components for smart manufacturing where Sensitive Information  is referred as Confidential data, such as manufacturing processes, cost information, operational data, intellectual property, and customer data~\cite{wong2017privacy}.
They also provide the requirements for secure data sharing as follows:
\begin{itemize}
\item Specifying a group of users or devices that are allowed to view the data by owner.
\item Any authorized member should be able to gain access to the data anytime and anywhere
\item Adding new users or devices to the permitted group
\item Revoking the access rights of any member or device of the group.
\end{itemize}
The smart factory is vulnerable to privacy leakage and information inference by internal and external attackers, since private information is collected, transmitted, and processed. The authors classify the data as following classes:
\begin{itemize}
\item \textbf{Sensitive}: have the most limited access and requires a high degree of integrity.
\item \textbf{Confidential}: less restrictive within the organization but might cause damage if disclosed.
\item \textbf{Private}: compartmental data that might not do the company damage but must be keep private for other reasons.
Proprietary: disclosed outside the company on a limited basis or contains information that
could reduce the company’s competitive advantage
\item \textbf{Public}: the least sensitive data used by the company and would cause the least harm if disclosed.\\
\end{itemize}
One trivial solution to achieve secure data sharing in IIoT requires the data owner to encrypt his or her data before sharing them with others; however, this approach requires additional computation power to decrypt the data before they can be used. In particular, the data owner needs to send the keys that are used for the data encryption to other parties; also, if the data owner revokes access rights to any user or device, he or she must re-encrypt the data with a new key, and distribute the new key to other parties in the group.\\

Additive Layer Manufacturing (ALM / AM) problem of Intellectual Property (IP) protection in the case of outsourcing. The existing process and introduce a new model for the outsourcing of ALM-based production. Because of both high costs of AML equipment and dependence for tuning of manufacturing parameters, outsourcing the production to third parties specializing in the ALM process is a must~\cite{yampolskiy2014intellectual}.
The outsourcing creates an additional attack surface that can be exploited. Intellectual Property (IP) includes 3D shape, required physical parameters owned by the 3D object designer, tuning of the manufacturing parameters usually owned by the ALM manufacturer. The IP of the ALM Manufacturer is protected by keeping it secret from the customer. It binds the customer to the particular ALM Manufacturer capable of tuning the manufacturing parameters according to the requirements.\\

The proposed model goes through the steps as: 3D Object Designer contacts the Tuning Expert and requests specifications of manufacturing parameters. The Tuning Experts provide the specification of the manufacturing parameters and the license to the 3D Object Designer. However, it posses some threats as increased number of actors and communication flows for the information exchange increase the probability for malicious actors to be involved. Multiple actors share the same information and identification of IP violators becomes difficult. The authors suggest an approach of Watermarks as they do not affect the required properties of the 3D object, are hard to remove from the object specification and can be used to uniquely identify all actors assuming different roles in the proposed outsourcing process.\\

\subsubsection{\textbf{Production}}
Authenticity of the supplied component is also doubtful if the contractor is not a reputed contractor or he/she did not follow the manufacturing standards.  A risk free and efficient product life cycle is required which minimizes the cybersecurity risks of the products and services. Cybersecurity risks may be defined as any abnormal activity like malicious behavior tainted by malicious actors, or products, services may be counterfeited or contain counterfeit circuits, components etc. which may be used for illicit purpose. There are various types of attacks as mentioned by Pal \textit{et al.}-
\begin{itemize}
\item Manufacturing backdoors may be created for malware or other penetrative purposes and may be embedded in radio-frequency identification (RFID) chips and memories.
\item Unauthorized access of protected memory. 
\item Inclusion of faults for causing the interruption in the normal behavior of the equipment.
\item Hardware tampering by performing various invasive operations
\end{itemize}
Through insertion of hidden methods, the normal authentication mechanism of the systems may  be bypassed. Hence, the efficient management of the supply chain of the cyber security products is the necessity.\\

Cloud users can request services ranging from product design, manufacturing, testing, management and all other stages of a product life cycle. a cloud manufacturing system framework, which consists of four layers, manufacturing resource layer, virtual service layer, global service layer and application layer. The main concern with cloud manufacturing for end-users is related to the storage of personal/enterprise sensitive data. This data includes not only product information but also information of some of the high-end manufacturing resources~\cite{xu2012cloud}. It aims to realize the full sharing and circulation, high utilization, and on-demand use of various manufacturing resources and capabilities by providing safe and reliable, high quality, cheap and on-demand used manufacturing services for the whole lifecycle of manufacturing. \\

\subsubsection{\textbf{Operation}}
Supply chain transparency can be achieved through horizontal integration across the supply chain, the compliance, control or the fulfillment of any other related corporate business process. In the process of factory floor, Radio Frequency Identification (RFID) or transponders in the material containers has bidirectional communication through interfaces. This may expose readings from sensors, alarms or reports or allowing recipes to be externally selected or downloaded by malicious intruder. RFID for the identification and tracking of products, packets, and pallets in supply chain scenarios is controlled by smart devices, such as smart phones and wearables smart watches with considerable computing capabilities and Internet connections~\cite{almada2016industry}.\\

For preventing any system failure that may result in physical damage or harm to humans. The integrity of Industrial IoT systems must be preserved which includes protection against sabotage~\cite{sadeghi2015security}. Sadeghi \textit{et al.} discussed about a key mechanism to verify integrity of a system’s software configuration as attestation. This is the process of detecting unintended and malicious software modifications to preserve integrity. One of the common approaches is remote attestation where the device to be attested, called prover, sends a status report of its current software configuration to another device, called verifier, to demonstrate that it is in a known and, thus trustworthy state. An efficient swarm attestation mechanism to collectively verify the software integrity of all devices.\\

The strong connectivity of IoT-based production systems and smart products demands for new mechanisms to protect against industrial espionage and privacy of customers and employees. Hence, the confidentiality of code, data, and configuration of production systems as well as blueprints of products is an important security requirement. In this regard, protection of design and configuration data (intellectual property) and detection of counterfeit components (product piracy) is a must. If a cyber attack occurs, affected IT systems are typically temporarily disabled and then restored after the attack. However, this approach cannot be applied to CPPS, where availability is a fundamental requirement.\\

Zhang \textit{et al.} proposes a cyber-physical system for patient-centric healthcare applications and services, called Health-CPS, built on cloud and big data analytics technologies. Electronic health records (EHRs), biomedical database, and public health have been enhanced not only on the availability and traceability but also on the liquidity of data through data fusion of EHR and electronic medical records (EMRs). Architecture of Health-CPS, which consists of three layers, namely, data collection layer, data management layer, and application service layer~\cite{zhang2017health}.\\

Data nodes can be divided into the following groups as Research data, Medical expense data, Clinical data, and Individual activity and emotion data. Based on the physiological data collected by wearable devices, the health status of a user can be easily monitored and traced. The individual emotion data are available to be collected through the information published on the social networks, which can be used in the mental health measuring and affective computing. The author proposed an Adapter being a middle-ware to provide a data node with access to the system, which is not only a physical data link but also a raw data preprocessor and encryption module. The encryption module encrypts the preprocessed data to ensure security through a hierarchical privacy preservation mechanism. Any unauthorized devices cannot decrypt the data package even if they have access to the system.\\

\subsection{Integrity}
\subsubsection{\textbf{Design}}
Zeltmann \textit{et al.} present a brief overview of the potential risks that exist in the cyber-physical environment of additive manufacturing (AM) evaluating the risks posed by two different classes of modifications to the AM process~\cite{zeltmann2016manufacturing}.The risks posed are examined through mechanical testing of objects with altered printing
orientation and fine internal defects. Finite element analysis (FEA) and ultrasonic inspection are also used to demonstrate the potential for decreased performance and for evading detection.
The vulnerabilities present in each stage that could affect the final AM product.
\begin{itemize}
\item Design: This phase includes computer-aided design (CAD) and finite element analysis (FEA). A CAD team models the product based on the desired dimensions, properties and functionalities.
\item Manufacturing: This phase includes slicing the 3D model and printing the object, and is where the AM process begins to diverge from traditional manufacturing. The final design of the object is converted to .STL format, which the slicer software then converts into a target-machine-specific tool path code; G-code.
\item Testing: For quality control or validation, a prototype printed part may be subjected to  mechanical and physical testing, In this scenario, a specimen is printed with a small defect embedded at the center.
\end{itemize}
From the security perspective, it can be hypothesized that an internal or external rouge person may be motivated to create defects in the 3D-printed parts with the aim of compromising the product performance. Alteration of the printing orientation can cause significant changes in the mechanical behavior which inspection of the part is also unable to detect. Many AM machines are always connected to the network for possible remote queuing of jobs, diagnostics, and monitoring. This connectivity creates vulnerabilities and opens up possibilities of attack from external parties. Large-scale jobs may be outsourced to commercial 3D printing services, which may not be entirely trustworthy. Malicious and deliberate modification of the print orientation or the introduction of defects are two possible attacks which may have devastating impact on the users of the final product.\\

AM is a layer-wise approach to fabrication, which provides opportunity to place defects internally, and its transformation of raw material in the process, which allows for altering material properties of the final product by changing the print parameters~\cite{sturm2014cyber}.
The open nature of the .STL file makes it easy to modify. The lack of volumetric information makes it more difficult for an attacker to determine where the most dangerous location to cause a defect is than if they were attacking the CAD file directly. By altering machine process parameters, changes to the material properties can be introduced that can be difficult to detect in the physical realm. The toolpath is the next most dangerous attack vector. Once an operator has validated the toolpath (if possible) the build is assumed to be correct. An attack that occurs after this point would have to be detected in the physical realm, either during printing or in final inspection. Such attacks could happen in the form of malware, wherein .STL files are intercepted and automatically altered without the operator’s knowledge.
\begin{itemize}
\item A traditional cyber-attack where the file is damaged or encrypted, this renders it inaccessible to the user where a corruption or encryption attack is a straightforward attempt to damage or extort the owner of the file.
\item Voids can be placed inside of a part and the material properties of internal layers can be changed without affecting the exterior layers, which makes detection with traditional part inspection techniques difficult.
\item Scaled up or down in one or more axes by simply applying a scaling factor to the vertex coordinates. Such an alteration would result in a changed form that may affect the fit or strength of the part.
\item Small protrusions or indents may be added to a part to affect the fit, surface finish, or strength of the part. This can be done by changing the coordinates of a single vertex, in this case making it identical to a vertex movement attack, or by adding and modifying facets and vertexes.
\item One or more vertices in the part is altered, resulting in a changed form that may affect the fit or strength.
\end{itemize}
To explore the potential implications of such an attack, a case study was conducted to evaluate the ability of human subjects to detect and diagnose a cyber-physical attack on the STL file of a test specimen. While in this study the void was inserted before the part was verified, the attack could also be designed to affect parts after the inspection stage instead. The authors recommend one solution as Hashing/secure signing/blockchain to ensure the validity of a file. The file is run into the hashing function, which generates a string of character called a hash. The hash is then posted along with the file. While AM pre-processing software are capable of detecting shells with negative volumes, this ability alone is not sufficient protection from a void .STL attack.\\

Straub \textit{et al.} discuss about the compromise of structural integrity by changing its position on a 3D printer’s build plate and attack scenarios where printing orientation changing can occur in the process~\cite{straub2017identifying}. An imaging-based solution is presented that will use a model that can be compared to the expected outcome of a printing. Positioning based attacks will typically require large-scale movement of the object to be effective. Digital object inspection involves the comparison of the produced object to the object that is expected to be produced including the differences in size, orientation, fill level, object design or printer configuration.\\

The solution approach requires the use of a different workstation or server to perform quality assurance processes and a secure intercommunication between the printer-control workstation and quality control workstation. Image data for the object-under-assessment is compared to data for a reference directly into a 3D printer unit taken from the same angle, Movement of an object will typically generate large and noticeable regions where filament is expected to be present.\\

Vincent \textit{et al.} discuss two case study for structural health monitoring process of 3d printing~\cite{vincent2015trojan}.
\begin{itemize}
\item \textbf{Case 1}: Designing a tensile test specimen using computer-aided design (CAD) software generate tool-paths using computer-aided manufacturing (CAM) software. The tool-path file was transferred to (CNC) milling machine. During this file transfer, a malicious software intercepted and altered the tool-path files.
\item \textbf{Case 2}: A malicious software was designed to modify STL files. An internal defect (void) was introduced within the part causing it to fail prematurely when tested. Malicious software analyzed the STL file to determine an optimal (with respect to causing the largest increase in stress concentrations) location to place the void.
QC approaches are not designed to detect the effects of cyber-attacks and they are based upon assumptions (sustained system shifts, rational sub-grouping, feature-based monitoring).
\end{itemize}
The proposed model use Side Channel analysis. Side-channels have used timing delays, leakage measurements, and temperature to build IC operational models. Through the careful selection of IC characteristics for model generation, attackers have little information on the side-channel measurements being used. Compromising a manufactured part can be done by simply adding an additional hole or changing the shape of a part, much like a Trojan can be introduced by simply adding an additional logic gate. This can result in a part performing sub-optimally or performing a different function entirely.\\

In AM process, kinetic cyber-attack can find its way through the digital process chain to introduce various inconspicuous flaws in the 3D objects. The authors introduce Kinetic Cyber-Attack Detection (KCAD) method using statistical modeling of the AM system to detect the anomalous analog emission~\cite{chhetri2016kcad}. The proposed approach goes through following models:
\begin{itemize}
\item Modeling of an Adversary to understand attack points in the digital process chain.
\item Statistical Estimation to model the behavior of the AM system by analyzing the relationship between the analog emissions and the control signals.
\item Analysis of Analog Emission to use it as a parameter from the side-channel using mutual information as the relation measurement metric.
\end{itemize}
The model function estimation can be done either online or offline as per the representation by Chhetri \textit{et al.} Offline function estimation can have shorter response time whereas online estimation can be done for higher accuracy with longer response time.Some limitations are found as, printer Variation since KCAD is machine specific and before implementation, the function estimation (training) has to be conducted before it can be implemented. In the experimental section, KCAD with firmware modification attacks introducing simple variation in the x, y axis. Moreover, analog emissions sensors placement should not obstruct the printing process.\\

\subsubsection{\textbf{Production}}
A scheme of verification and intrusion detection that is independent of the printer firmware and controller PC. The scheme incorporates analyses of the acoustic signature of a manufacturing process, real-time tracking of machine components, and post production materials analysis.\\

Bayens \textit{et al.} proposes the use of the acoustic signal, embedded materials, and spatial position of machine components~\cite{bayens2017see}. The first two layers are achieved through acoustic side-channel analysis and spatial sensing which analyze the sound and physical position of printing components respectively. The third layer is that of materials verification in which imaging techniques are used to verify that the print is made from the proper material. The solution approach use an audio classification scheme similar to popular apps used for identifying music understanding the process without relying on control software. Acoustic signals have been explored as a method of understanding information being processed by both traditional printers and a training set by recording it with a microphone to obtain an audio file.\\

The classification results are such that the index values appear in ascending order. The confidence score of one or more indexed classification results falls below a given threshold value. Its value is optimized manually for each printer to maximize the true positive rate and minimize the false positive rate to describe the identification of a malicious print, the observation of the detected error, and the post production materials verification.
Some limitations are found as:
\begin{itemize}
\item Detecting a deviation from a training print decreases as the similarity to the print increases.
\item A training print is needed.
\item Mass production scheme detection is time consuming. 
\item If a third party printing service implements these methods, some cost overhead will incur from the purchase of microphones, sensors.
\end{itemize}

Moore \textit{et al.} discuss an approach of continuous monitoring of the current/electricity delivered to all actuators during the manufacturing process and detection of deviations~\cite{moore2017power}. It provides the advantages over a time-critical process and being air-gapped from the computerized components involved in the AM process. The solution method described as the firmware communicates with the on-board motor controllers, which deliver current at a given frequency and amplitude to actuate the motors. Since a set of delivered current over time for each motor will always result in the same printed object, Attack detection method based on the comparison of motor current traces:
\begin{itemize}
\item Generating traces from several known-good prints
\item Collecting the current traces of subsequent prints of the same
object on that printer
\item Comparing the captured traces against the known good traces 
While various factors, including the mechanical arrangement of the motors, the filament, and the temperature of the extruder and bed, may influence the translation from current to physical object any analog representation is not alterable by cyber means.
\end{itemize}

All devices whether industrial machines, computer, tablets, or smart phones needs to be updated on regular basis whether to avoid threats or due to configuration changes installed in these devices spread across the geographical location or inside factory~\cite{khan2016survey}.\\

\subsubsection{\textbf{Operation}}
A discussion about implementation of malicious code using Printrbot’s branch of the open source Marlin 3D printer firmware~\cite{moore2017implications}. The solution methodology follows:
\begin{itemize}
\item Investigate the control flow and format of commands handled by the firmware.
\item The source code for Printrbot’s Marlin Firmware was downloaded, compiled and flashed to the 3D printer.
\item Upon validation of the newly flashed firmware, a further review of the communication protocol was performed.
\item Incoming commands are transferred via USB or through an onboard SD card.
\item To determine the command process flow, a manual code review was performed tracing the path of incoming G-code.
\item To validate the format used for data transfer USB packet capture was performed using USBPcap with Wireshark during a test print.
\end{itemize}

\subsection{Availability}

\subsubsection{\textbf{Design}}
Industry 4.0  introduces an innovative business models and user experiences through strong connectivity along with effective use of next generation smart equipment. These smart systems generate, process, and exchange vast amounts of security-critical and privacy-sensitive data and due to these characteristics, they become attractive targets of attacks~\cite{sadeghi2015security}. In smart factories, smart products know their own identity, history, specification, documentation, and even control their own production process. Smart products do not only collect data during their production but also when they are deployed and used by customers.The most important objective of industrial production systems is availability, which should prevent any unnecessary delay in production that results in loss of productivity and loss of revenues.\\

\subsubsection{\textbf{Production}}
Khan \textit{et al.} introduce some aspects of manufacturing security challenges in the following concerns~\cite{khan2016survey}. Usage of IoT devices contribute largely in the amount, heterogeneity, and speed of the data generated at the production level. Such data poses various challenges and demands new methodologies for storing, processing, and management. Redundant data is stored in various departments of the company, in different data formats with minor extensions or enrichments. Such data silos raise the amount of data redundancy, inconsistency, and different interpretation of data. So the challenge is availability of the needed data for analysis purpose. Since, real time information access for processes was not available at shop floor level, so in case of change in processes or actions, workers or machines have to wait until instructions are manually transferred or data is loaded in the production system. Data collected from machines and business processes is filtered, analyzed, and then delivered in required format to provide insights which in return will help to give better process control, optimize, and reduce overhead costs.\\

\subsubsection{\textbf{Operation}}
The supply chains being highly transparent and integrated, physical flows will be continuously mapped on digital platforms which will CPPS available to accomplish the needed activities to create each tailored product. Pal \textit{et al.} introduced factory-floor is the marketplace of capacity and production, where smart materials and smart equipment cooperate autonomously. If the system is decentralized systems, it will always need a centralized system for compliance, optimization and monitoring.\\

Advanced analytics are then needed to fully understand the performance of the manufacturing processes, quality of products and supply chain optimization with proper security measurements~\cite{pal2017cyber}.\\

\section{Challenges in Smart Manufacturing}
\label{sec:CSM}
Smart manufacturing being a widely technological and connected system needs to perform continuously to provide proper operational facility in the industrial production. From the literature survey it can be concluded that the CIA should be maintained for a secure operation. Though cybersecurity is important for each aspect however, for industrial purpose 'integrity' protection is the challenging phenomena. A simple change in the production parameter will cause a severe failure in the economic condition of a factory.External events like hazards deal with risks owing to events, both natural and of human origin, over which the organization has no control and that can occur without notice. The elements supporting this subclass include weather event, fire, flood, earthquake, unrest, and pandemic. During this calamities malicious people can take advantage of any power outage or network loophole to get into the system to cause harm to a system. Protecting the integrity also coupled with the continuous operation or availability. Since the emerging technology is building its path to distributed manufacturing, securing \textit{data-over-net} concept will be a new challenge for the manufacturing thread. Proper encryption and cryptographic solution need to be deployed for secure communication in this regard and maintain the confidentiality of the system process. As per the the discussion above, the solution approaches by several authors show the toughness of proper detection of any cyber attack due to the sophisticated, widely connected manner of industrial operation. Hence, the main challenge is to determine the threat model in a faster approach to mitigate and preserve the \textit{defense-in-depth} of cyber space.


\section{Conclusion}
\label{sec:conclusion}
As manufacturers demand the greater connectivity and interaction of the fourth industrial revolution, manufacturing-system suppliers, machines and systems in their factories will have to expand the role of IT in their products. With the expansion the obvious question arises whether the system, in this concern, process is secure or not. From the literature review, it is considerable that the cyber attachment of the smart factories will be vulnerable to security breach and intrusion. Any small disruption or compromise in the system operation will result into mass production failure and safety critical situations. To overcome the challenges in the existing new era of smart industry, security assessment will play a vital role for regular continuation of manufacturing chain cyber domain. Awareness among the active entities will also be required to understand the vulnerabilities. Since digital thread comprises of multiple domains, each of them can be vulnerable and suitable assessment methodology should be there on the level of the operational technology. As for our future work, a brief focus on the Information Technology domain including IoT, cloud manufacturing and distributed industrial operation will be assessed. 


\balance
\bibliographystyle{unsrt}
\bibliography{References}

\end{document}